\documentclass{tufte-handout}

\usepackage[utf8]{inputenc}
\usepackage{amsmath}
\usepackage{amsfonts}
\usepackage{amssymb}
\usepackage{txfonts}
\usepackage{listings}

\usepackage{hyperref}
\usepackage{cleveref}

\usepackage{graphicx}
\usepackage{caption}
\usepackage{subcaption}
\usepackage{colorblind}

\usepackage{tikz}
\usetikzlibrary{positioning}

\tikzset{
    node/.style={circle, draw, minimum size=0.5cm},
    causaltriangle/.pic={
      \tikzset{
        node distance=8mm and 8mm
      }

      \node[node, fill=gray] (z) {$z$};
      \node[node, fill=red, below left=of z] (x) {$x$};
      \node[node, fill=cyan, below right=of z] (y) {$y$};
  }
}

\title{Tutorial Debriefing: Applied Statistical Causal Inference in Requirements Engineering}
\author{Julian Frattini, Hans-Martin Heyn, Robert Feldt, Richard Torkar}

\begin{document}

\maketitle

\newthought{As any scientific discipline}, the software engineering (SE) research community strives to contribute to the betterment of the target population of our research: software producers and consumers.
We will only achieve this betterment if we manage to transfer the knowledge acquired during research into practice.
This transferal of knowledge may come in the form of tools, processes, and guidelines for software developers.
However, the value of these contributions hinges on the assumption that applying them \emph{causes} an improvement of the development process, user experience, or other performance metrics.
Such a promise requires \emph{evidence} of causal relationships between an exposure or intervention (i.e., the contributed tool, process or guideline) and an outcome (i.e., performance metrics).
A straight-forward approach to obtaining this evidence is via controlled experiments in which a sample of a population is randomly divided into a group exposed to the new tool, process, or guideline, and a control group. 
However, such \emph{randomized control trials} may not be legally, ethically, or logistically feasible. 
In these cases, we need a reliable process for statistical causal inference (SCI) from observational data.

\newthought{Much will be won} if researchers in SE adopt one of the simplest techniques in scope of SCI: causal modeling. 
We can visualize causal assumptions in the form of directed acyclic graphs (DAGs), where nodes represent variables and directed edges represent assumed causal relationships.\cite{glymour2008causal}
Vice versa, the absence of an edge represents the assumptions that two variables are not directly related.
The figure below shows a simple DAG representing the causal assumption that AI-support affects coding performance.

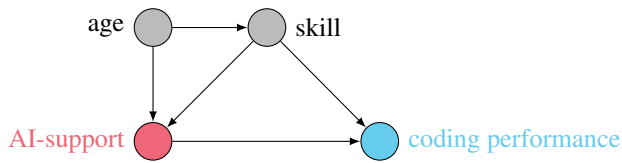
\begin{figure}
    \centering
    \begin{tikzpicture}[>=latex, every path/.style={->}]
    	\node[node, fill=gray, label=left:{age}] (age)   {};
		\node[node, fill=gray, right=of age, label=right:{skill}] (skill) {};
		\node[node, fill=red, below=of age, label=left:{\textcolor{red}{AI-support}}] (ai) {};
		\node[node, fill=cyan, right=of ai, xshift=15mm, label=right:{\textcolor{cyan}{coding performance}}] (code) {};

		\draw (age) -- (skill);
		\draw (age) -- (ai);
		\draw (skill) -- (ai);
		\draw (skill) -- (code);
		\draw (ai) -- (code);
    \end{tikzpicture}
\end{figure}

Additionally, the figure encodes the assumption that the \emph{skill} of a developer both affects their inclination to use AI-support and their overall coding performance.
Similarly, \emph{age} is assumed to affect skill but also the aforementioned inclination.

Do you agree with this DAG?
You might not.
Based on your assumptions, experience, or prior research, you may find additional variables relevant or you would connect them differently.
A DAG always represents a researcher's \emph{understanding} of real-world phenomena.
But a DAG makes this understanding transparent and captures even complex relationships that may not be easily expressed in a null-hypothesis stated in natural language.
Its transparency allows to review, critique, and improve it by proposing the inclusion of new or exclusion of existing variables, and updating the edges between them.\footnote{Try to think of other variables that you believe or know would affect the effect of AI-support on coding performance.}
In fact, DAGs \emph{encourage} challenging them:
SCI has methods to determine which one of several potential DAGs reflects the real-world phenomenon better.
This way, we can incrementally improve our understanding of these phenomena.

\newthought{Moreover, DAGs expose potential sources of bias}.
In observational studies, the effect of a treatment of interest (here: AI-support) on an outcome of interest (here: coding performance) may be biased by other variables.\footnote{This does not happen in true controlled experiments where the treatment is assigned at random. In a DAG, this would mean that no arrow ``enters'' (i.e., points towards) the treatment.}
But when you understand how other factors bias this effect of interest, you can take counter measures.
There are essentially three different relationships in which a third variable $z$ can interact with an effect of interest of $x\rightarrow y$, visualized as the three DAGs below.

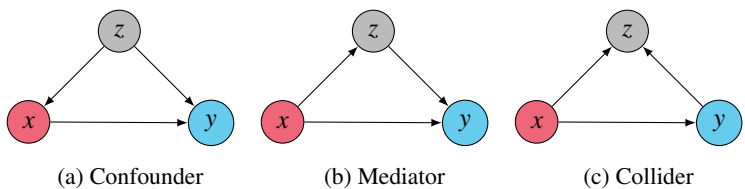
\begin{figure}
    \centering
    \begin{subfigure}[t]{0.3\textwidth}
        \begin{tikzpicture}[>=latex, every path/.style={->}]
  			\pic (T) {causaltriangle};
			\draw (Tz) -- (Tx);
			\draw (Tz) -- (Ty);
			\draw (Tx) -- (Ty);
		\end{tikzpicture}
        \caption{Confounder}
    \end{subfigure}
    \begin{subfigure}[t]{0.3\textwidth}
        \begin{tikzpicture}[>=latex, every path/.style={->}]
  			\pic (T) {causaltriangle};
			\draw (Tx) -- (Tz);
			\draw (Tx) -- (Ty);
			\draw (Tz) -- (Ty);
		\end{tikzpicture}
        \caption{Mediator}
    \end{subfigure}
    \begin{subfigure}[t]{0.3\textwidth}
        \begin{tikzpicture}[>=latex, every path/.style={->}]
  			\pic (T) {causaltriangle};
			\draw (Tx) -- (Tz);
			\draw (Tx) -- (Ty);
			\draw (Ty) -- (Tz);
		\end{tikzpicture}
        \caption{Collider}
    \end{subfigure}
\end{figure}

These three different types of relationship require different treatment:

\begin{enumerate}
    \item \textbf{Confounder}: If $z$ is a common cause of both $x$ and $y$ ($x \leftarrow z \rightarrow y$) we \emph{must} control for it statistically\footnote{``Controlling for'' means including the variable in a statistical analysis, for example as a regressor in a linear model.} when estimating the effect $x \rightarrow y$. Otherwise, $z$ will bias this effect. This is the classic case of confounding.
    \item \textbf{Mediator}: If $z$ is located on a pipe\footnote{A pipe is a causal chain where information flows sequentially from one variable to the next.} from $x$ to $y$ ($x \rightarrow z \rightarrow y$) we \textit{can} control for it. If we do control for $z$, we distinguish the direct effect ($x \rightarrow y$) from the indirect effect through $z$ ($x \rightarrow z \rightarrow y$). If we do not control for $z$, we obtain the total effect (i.e., combining the direct ($x \rightarrow y$) with the indirect ($x \rightarrow z \rightarrow y$) effect). Both are correct, but answer different questions.
    \item \textbf{Collider}: If $z$ is a common effect of both $x$ and $y$ ($x \rightarrow z \leftarrow y$) we \emph{must not} control for it. By default, the effect $x \rightarrow y$ is unbiased in this case, but controlling for $z$ will introduce a bias.\footnote{You may know this phenomenon as Berkson's paradox.}\cite{berkson1946limitations}
\end{enumerate}

Handling these relationships is a major aspect of research design.
Controlling confounders ensures internal validity, while ignoring colliders makes sure one does not condition on post-treatment variables.
Controlling mediators allows isolating the direct effect of a treatment.
For example: does a newly proposed testing technique $x$ really improve test coverage $y$ (i.e., the direct effect $x \rightarrow y$) or does it rather affect testers' attention $z$ which in turn produces tests with better coverage (i.e., the indirect effect $x \rightarrow z \rightarrow y$)?
Depending on which effect is stronger, a study may conclude either to employ the testing technique or heighten the attention of testers.

\newthought{Understanding potential sources of bias} enables debiasing a causal effect of interest even when dealing with data from an observational study.
The following framework proposed by Julien Siebert\cite{siebert2023applications} (derived from the seminal work of Judea Pearl\cite{pearl2018book}) includes three steps for SCI:

\begin{enumerate}
    \itemsep0em
    \item \textbf{Modeling}: Formalize your causal assumptions in a DAG. 
    \item \textbf{Identification}: Given the relationships between variables, identify those that you need to statistically control for in order to de-bias the causal effect of interest.
    \item \textbf{Estimation}: Estimate the causal effect of interest with statistical means (e.g., regression models) while controlling for the previously identified variables.
\end{enumerate}

\newthought{Adherence to SCI principles} will propel our scientific endeavors in SE research beyond limitations of correlational analyses and the constraints of experimentation.
Beyond the ability to de-bias a causal effect of interest, SCI principles contribute further benefits to a scientific discipline:

\begin{itemize}
    \itemsep0em
    \item The transparency of causal assumptions facilitates iterative improvement of empirical evidence. Reviewing, improving, and comparing causal DAGs maps individual pieces of evidence to the continuous improvement of our understanding of SE phenomena.
    \item The identification of variables potentially biasing an effect of interest informs data collection, i.e., which factors $z$ to record in addition to $x$ and $y$ such that we can control for them in the analysis.
    \item Causal DAGs facilitate a clearer and more honest discussion of threats to validity. Instead of relying on common practice, threats to internal validity can be connected to variables that bias the effect of interest but could not be collected.
\end{itemize}

\newthought{This is a complex topic} one should not expect to master in a day.
Proper SCI is much more powerful but also takes much more effort than just picking an appropriate null-hypothesis significance test.
We hope the tutorial and\slash or this debriefing inspired you to get into SCI for SE.
In case of questions, we are more than happy to support you in learning more about it.

\section{Resources}

Finally, find some useful resources to consult when undertaking the journey of learning SCI.
The publications of many great authors in the recent years have made SCI fairly accessible to the interested researcher:

\begin{itemize}
    \item ``The Book of Why'' by Judea Pearl\cite{pearl2018book} is written for a general audience and masterfully introduces the fundamental ideas, history, and methods of SCI.
    \item ``Statistical Rethinking'' by Richard McElreath\cite{mcelreath2018statistical} is the extensive text book teaching the craft of SCI in minute detail. On top of being grounded in causal principles, the author adopts a Bayesian perspective to data analysis.
    \item ``A Crash Course in Good and Bad Controls'' by Carlos Cinelli and colleagues\cite{cinelli2024crash} elaborates which variables to control for given a causal DAG. Thanks to several illustrative examples, this is a very accessible and rewarding read.
    \item ``Applications of Statistical Causal Inference in Software Engineering'' by Julien Siebert\cite{siebert2023applications} summarizes prior work using SCI in SE. This literature study shows that its application is yet limited - meaning that it is the perfect time to get started and join the frontier of SCI in SE.
    \item ``Applying Bayesian Analysis Guidelines to Empirical Software Engineering Data'' by Carlo Furia and colleagues\cite{furia2022applying} applies SCI principles in SE research. This is one of the best practical demonstrations of SCI for our field.
    \item ``Causal Models in Requirement Specifications for Machine Learning: A vision'' by Hans-Martin Heyn and colleagues\cite{heyn2025causal} goes even further and applies SCI principles not only to SE research, but to SE practice as well, underlining their overall potential.
\end{itemize}

Many more great references could be mentioned here, and we hope that many more great references will be added to the list in the future to help our requirements and software engineering community adopt methods of SCI in our portfolio.
For actual application and tooling, we recommend the programming language \texttt{R}\footnote{\url{https://www.r-project.org/}} and the following packages:

\begin{itemize}
    \itemsep0em
    \item To draw and evaluate DAGs, we can recommend DAGitty\footnote{\url{https://www.dagitty.net/}} or GGDag.\footnote{\url{https://r-causal.github.io/ggdag/}} 
    \item To estimate causal effects based on DAGs using a Bayesian perspective, consider following along the above-referenced book ``Statistical Rethinking'' using the \texttt{rethinking} package.\footnote{\url{https://github.com/rmcelreath/rethinking}} Once having understood the principles, we can recommend switching to \texttt{brms} by Paul B{\"u}rkner.\footnote{\url{https://cran.r-project.org/web/packages/brms/index.html}}
\end{itemize}

The material from our tutorial remains available under an open source license on GitHub\footnote{\url{https://github.com/JulianFrattini/bda4sci}} and Zenodo.\footnote{\url{https://doi.org/10.5281/zenodo.17649808}}
It includes the presentation slides as well as code examples to recreate figures and analyses.

\section{History}

Finally, we summarize the history of the tutorial series that aims to bring SCI knowledge to the SE community.
Spiritual predecessors were a similar tutorial held by Richard Torkar, Carlo Furia, and Robert Feldt at ICSE'21\cite{torkar2021bayesian} and by Robert Feldt during ESEM'22.\footnote{\href{https://conf.researchr.org/details/esem-2022/esem-2022-eseiw-isern/3/Session-4-Modernizing-Data-Analytical-Methods-in-Empirical-SE}{https://conf.researchr.org/details/esem-2022/esem-2022-eseiw-isern/}}
Based on this prior work, we established the following tutorial series:

\begin{enumerate}
    \item At the 33\textsuperscript{rd} IEEE International Requirements Engineering conference we presented the first version of this tutorial as a joint effort from the four authors.\footnote{\href{https://conf.researchr.org/track/RE-2025/RE-2025-tutorials\#tutorial-3-applied-statistical-causal-inference-in-requirements-engineering}{https://conf.researchr.org/track/RE-2025/RE-2025-tutorials}} The 3h-session was attended by 10-20 participants of different levels of seniority, resulting in a lively discussion about SCI principles and their applicability to SE/RE. Causal modeling and identification with DAGs was quickly adopted, but the use of simulated data to prove the correctness of statistical models was identified as rather ill-established in SE/RE.
\end{enumerate}

Future updates will be documented in an revised version of this document.
The beautiful plots were created by Michael Dorner~\footnote{\url{https://www.michaeldorner.de/}} using \texttt{tikz}.

\nobibliography{references}
\bibliographystyle{abbrvnat}

\end{document}